\documentclass[final,authoryear,5p]{elsarticle}

\usepackage{epsfig}

\usepackage{amssymb}

\usepackage[ps2pdf,%
a4paper=true,%
breaklinks=true,%
colorlinks=true,%
pdfauthor={First Author et al.},%
pdftitle={Template for manuscripts in Advances in Space Research}%
]{hyperref}

\journal{Advances in Space Research}

\begin{document}

\begin{frontmatter}



\title{The OIV 1407.3\AA /1401.1\AA\ emission-line ratio in a plasma}


\author{Nabil Ben Nessib}
\address{Department of Physics and Astronomy, College of Science, King Saud University. PO Box 2455,\\
 Riyadh 11451, Saudi Arabia. \\}
\cortext[cor]{Corresponding author}
\ead{nbennessib@ksu.edu.sa}


\author{Norah Alonizan}
\address{Department of Physics and Astronomy, College of Science, King Saud University. PO Box 2455,\\
 Riyadh 11451, Saudi Arabia. \\}
\ead{nalonizan@ksu.edu.sa}

\author{Rabia Qindeel}
\address{Department of Physics and Astronomy, College of Science, King Saud University. PO Box 2455,\\
 Riyadh 11451, Saudi Arabia. \\}
\ead{rqindeel@ksu.edu.sa}

\author{Sylvie Sahal-Br\'{e}chot}
\address{Laboratoire d'\'{E}tude du Rayonnement et de la
Mati\`{e}re en Astrophysique,Observatoire de Paris,\\
UMR CNRS 8112, UPMC, 5 Place Jules Janssen, 92195 Meudon Cedex, France.\\}
\ead{Sylvie.Sahal-Brechot@obspm.fr}

\author{Milan S. Dimitrijevi\'{c} \corref{cor}}
\address{Astronomical Observatory, Volgina 7, 11060 Belgrade,
Serbia.\\Laboratoire d'\'{E}tude du Rayonnement et de la
Mati\`{e}re en Astrophysique,Observatoire de Paris,\\
UMR CNRS 8112, UPMC, 5 Place Jules Janssen, 92195 Meudon Cedex, France.\\}
\ead{mdimitrijevic@aob.bg.ac.rs}

\begin{abstract}

Line ratio of O IV 1407.3 \AA/1401.1 \AA\- is calculated using mostly our own atomic and collisional data.

Energy levels and oscillator strengths needed for this calculation have been calculated using a
Hartree-Fock relativistic (HFR) approach. The electron collision strengths introduced in the statistic equilibrium equations
are fitted by polynomials for different energies.
Comparison has also been made with available theoretical results.

The provided line ratio has been obtained for a set of electron densities
from  $10^{8}$ cm$^{-3}$ to $10^{13}$ cm$^{-3}$ and for a fixed temperature of 50 000 K.

\end{abstract}

\begin{keyword}
atomic data; atomic processes; line ratio
\end{keyword}

\end{frontmatter}

\parindent=0.5 cm

\section{Introduction}
\label{Section 1}

Triply charged oxygen ion (O IV) belongs to the boron isoelectronic
sequence, its ground state configuration is 2s$^{2}$ 2p $^{2}$P$^{o}$.
The boron like ions are widely observed in a variety of astrophysical plasmas.
Many researchers have been done a lot of calculations for many parameters of O IV due to its importance in astrophysical plasma. \cite{Aggarwal08}
calculated energy levels, radiative rates (A values) and excitation rates or equivalently the effective collision strengths (gamma) which are obtained from the electron impact collision strength ($\Omega$).
 \cite{Feldman97} presented a solar coronal spectrum recorded by the extreme UV spectrometer SUMER on the Solar
and Heliospheric Observatory. The detected O IV lines covered the wavelength range 500-1610 \AA. The studied range contains the
 2s$^{2}$ 2p $^{2}$P$^{o}$ -2s 2p$^{2}$ $^{4}$P intercombination transitions. \cite{Harper99} obtained a high signal-to-noise ratio spectra of RR Tel at medium resolution with the Goddard High-Resolution Spectrograph (GHRS) on the Hubble Space Telescope (HST) to test
available atomic data for the intercombination transitions in O IV (multiplet: UV 0.01). \cite{Blair91} observed the far-ultraviolet spectrum of the Cygnus Loop supernova remnant using the Hopkins Ultraviolet Telescope aboard the Astro-1 space shuttle. Similarly, \cite{Redfield02} have been detected lines of O IV in late-type stars within the wavelength range of 910-1180 \AA.
\cite{Sturm02} studied emission lines of O IV in IR range from active galactic nuclei and \cite{Pagano00} observed them in the solar transition region.
 \cite{Keenan09} generated theoretical UV and extreme-UV emission line ratio for O IV and presented their strong versatility for diagnostics of electron temperature and electron density for astrophysical plasmas.
Extreme-ultraviolet lines emission arising from radiative transition probabilities and electron collision cross-sections in B-like ions were calculated by \cite{Flower75}.

Using semiclassical approach, \cite{Milan95} calculated electron-, proton-, and He III-impact line widths and shifts for 5 multiplets of O IV.
The widths, shifts and fitting coefficients data are in the database STARK-B \citep{Starkb}.

Recently, \cite{Olluri13} studied the non-equilibrium ionization effects on the line ratio of O IV.

In this work, we will study the influence of the atomic data on the emission line ratio I(1407.1 \AA )/I(1401.1 \AA ).

\section[]{Atomic data}
\label{Section 2}

In this work, energy levels and oscillator strengths are calculated with the Hartree-Fock relativistic (HFR)
approach using Cowan code \citep{Cowan81} and an atomic model including the 13-configurations: 2s$^{2}$ 2p, 2s 2p$^{2}$, 2p$^{3}$, 2s$^{2}$ 3s, 2s$^{2}$ 3p, 2s$^{2}$ 3d, 2s$^{2}$ 4s, 2s$^{2}$ 4p, 2s$^{2}$ 4d, 2s$^{2}$ 4f, 2s 2p 3s, 2s 2p 3p and 2s 2p 3d.

The obtained energy level data are compared in Table 1 with those of  NIST \citep{NIST}, GRASP2 \citep{Aggarwal08} (using 219 levels) and MCHF \citep{Tayal06}.

The spontaneous emission radiative rate (the so-called Einstein's A-coefficient), $A_{ji}$ is related with the absorption oscillator strength $f_{ij}$ for the transition $i\rightarrow j$ by the standard relation:

\begin{equation}
A_{ji} ({\rm a.u}) = \frac{1}{2}\alpha ^3 \frac{{g_i }}{{g_j }}E_{ji}^2 f_{ij}
\end{equation}
where $E_{ji}$ is the transition energy between states $i$ and $j$ in Rydberg units, $\alpha$ is the fine structure constant and $g_i$ and $g_j$ are the statistical weight factors of the initial and final states respectively.

The transition probability in s$^{-1}$ is then:

\begin{equation}
A_{ji} ({\rm s}^{ - 1} ) = \frac{{A_{ji} ({\rm a.u})}}{{\tau _0 }}
\end{equation}
with $\tau _0 $ = 2.4191 $\times$ $10^{-17}$ is the atomic unit of time.

In Table 2, we have compared our values of log($g_{i}f_{ij}$) and transition probability $A_{ji} ($s$^{ - 1} )$ with the values taken from NIST database \citep{NIST} and results from \cite{Galavis98}, \cite{Flower75} and \cite{Tayal06}. Observed wavelengths are from \cite{Bromander69}.

\section{Collisional data}
\label{Section 3}

The excitation rate coefficient in cm$^3$s$^{-1}$ can be expressed by the effective collision strength as:

\begin{equation}
C_{ij}  = \frac{{8.629 \times 10^{ - 6} }}{{g_i T^{1/2} }}\gamma _{ij} (T)\exp \left( { - \frac{{E_j  - E_i }}{{kT}}} \right)
\end{equation}
where $\gamma _{ij} (T)$ is the effective collision strength related to the collision strength $\Omega _{ij}$ by:

\begin{equation}
\gamma _{ij} (T) = \int_0^\infty  {\Omega _{ij} \exp \left( { - \frac{{E_j }}{{kT}}} \right)} d\left( {\frac{{E_j }}{{kT}}} \right)
\end{equation}

The excitation rate coefficient $C_{ij}$ can also be written as a function of the cross-section:

\begin{equation}
C_{ij}  = N_e \int_{v_0 }^\infty  {v\sigma _{ij}(v)f(v)dv}
\end{equation}
where $v_0$ is the threshold velocity and $f(v)$ is the Maxwellian velocity
distribution function for electrons.

The semiclassical cross-section $\sigma _{ij}(\upsilon )$ can be expressed by an integral over the impact
parameter $\rho \ $ of the transition probability $P_{ij}(\rho ,\upsilon )\ $ as

\begin{equation}
\sigma _{ij}(\upsilon
)=\frac{1}{2}\pi R_{1}^{2}+\int_{R_{1}}^{R_{D}}2\pi \rho d\rho
P_{ij}(\rho ,\upsilon ).
\end{equation}
$R_{D}$ is the Debye radius and $R_{1}$ is the minimum cut-off radius (see \citet{Sylvie69}, and \citet{Seaton62} for neutrals).

\section{Line ratio}
\label{Section 4}

In a steady state regime, we can write that:



\begin{eqnarray*}
\frac{{d{N_i}}}{{dt}} = {\sum\limits_{j \ne i,j > i} {{N_j}\left( {{C_{ji}} + {A_{ji}}} \right)
 + \sum\limits_{j \ne i,\,j < i} {{N_j}\,{C_{ji}}} } }  \\
 \;\;\;\;\;\;\;- {N_i}\left[ {\sum\limits_{j \ne i,\;i < j} {{C_{ij}}}  + } \right.\left. {\sum\limits_{j \ne i,\;i > j} {\left( {{C_{ij}} + {A_{ij}}} \right)} } \right] \\
\end{eqnarray*}
\begin{equation}
\;\;\;\;\;\;\;\;=0,
\end{equation}
where $N_i$ is the number of ions per unit of volume for the energy level $i$, $C_{ij}$ is the electron collisional rate for transferring ions from level $i$ to level $j$, and $C_{ji}$ is the collisional rate for transferring ions from level $j$ to level $i$. In fact, since the local incident radiation is very weak, absorption and induced emission rates are completely negligible compared to collisional rates. So, for transferring ions from level $i$ to level $j$ (or from level $j$ to level $i$)  by radiative processes, only spontaneous emission rates $A_{ij}$ (or $A_{ji}$) are to be taken into account.

Solving the above set of equations, we obtain the relative populations $N_i/N_j$ for any two levels. The normalization is obtained by the constraint $\sum\limits_iN_i=N_{ion}$, where $N_{ion}$ is the ion density per unit volume.

The local emissivity in the line studied (in units of energy, per unit of volume, per unit of time, and per steradian) for a transition from state $j$ to state $i$ is given by:

\begin{equation}
\varepsilon_{ji}  = \frac{{h\nu }}{{4\pi }}A_{ji} N_j
\end{equation}
where $h$ is the Planck constant, and $\nu$ the frequency of the transition.

In ``mode 1" of \cite{Dufton77}, the observed intensity ratio of two lines is equal to the relevant line emissivity ratio. This is currently assumed in UV coronal studies, and justified when the two lines are formed in the same region of the plasma, and in addition if the emissivities vary slowly in the regions of emission. As \cite{Keenan09}, we will assume that it is the case for the considered O IV lines.

\cite{Keenan09} studied the line ratio I(1407.3 \AA )/I(1401.1 \AA ) versus electron density for constant electron temperatures.
They used the \cite{NIST} energy values in their calculations. The radiative transition probabilities are taken from different sources cited in their paper and from \cite{Aggarwal08} for the excitation rates.

In our present work, we have used our own calculated energies and oscillator strengths (HFR calculations) that give more self-consistent data. For the collision strengths, we have interpolated linearly the 3 collision strengths given by \cite{Flower75} for the energies 2.8, 4.2 and 5.6 Ry.

Then our atomic energy levels, oscillator strengths and interpolation of the collision strengths have been introduced in the statistical equilibrium code of \cite{Dufton77} that we have modified to give the intensities calculated both as photons emitted and as energy emitted versus electron density for a given electron temperature.

We have not taken into account the collisions with protons. In fact, our purpose is to study the effect on the behavior of the line ratio using different atomic data from different sources.

Putting different energy levels from different sources have not changed the behavior of the curve but when
different oscillator strengths have been used, there is a change. Figure \ref{figure1} shows the line ratio that we obtain when using the oscillator strengths from \cite{Galavis98}  (solid curve) and when putting our HFR data (dashed curve). The difference is increasing with the electron density to reach 14\%  for ${\rm{log}}(Ne)=13$.

\section{Conclusion}
We generally used our own energy levels and oscillator strengths for calculating Stark widths and shifts of spectral lines in our previous articles. In the present work, we have introduced cross sections, which are a part of the Stark impact width in the statistical equilibrium in a plasma to determine line ratio that can be an important diagnostic for studying plasma. We have tested the influence of using different sources of atomic and collisional data.
We have used a powerful method to calculate the line ratio I(1407.3 \AA )/I(1401.1 \AA ) by introducing homogenous atomic and collisional data that are mostly calculated by us.

Tests for other lines and for other ions will be done as well as the study of the influence of different parameters (atomic collisional data and parameters of the plasma).

\section*{Acknowledgments}

This project was supported by King Saud University, Deanship of Scientific Research, College of Science Research Center. The support of Ministry of Education, Science and Technological Development of Republic of Serbia through projects 176002 and III44002 is acknowledged.
The support of the `Programme National de Physique Stellaire" (PNPS, INSU-CNRS) and of the LABEX Plas@Par (UPMC, PRES Sorbonne Universities, Paris, France) are also acknowledged.

\begin{table}
  \centering
  \caption{Comparison of our present calculation of energy levels (E$_{Present}$) with those of NIST \citep{NIST}, GRASP2 \citep{Aggarwal08} (using 219 levels) and MCHF \citep{Tayal06}. Energies are in Ry.}\label{Tab1}
  \resizebox{8cm}{!} {
  \begin{tabular}{cllccccc}
    \hline
Index	& Configuration &Level			    & $J$	    & E$_{Present}$    & E$_{NIST}$	  & E$_{GRASP2}$	  & E$_{MCHF}$		  \\
\hline										
1	&2s$^{2}$ 2p	    ($^{1}$S)	&$^{2}$P$^{o}$	&0.5	&0.00000	   &0.00000		  &0.00000		  &0.00000		  \\
2	&2s$^{2}$ 2p	    ($^{1}$S)	&$^{2}$P$^{o}$	&1.5	&0.00347	   &0.00352	  &0.00339	  &0.00337	  \\
3	&2s  2p$^{2}$	($^{3}$P$^{o}$)	&$^{4}$P	&0.5	&0.64455	   &0.65101	  &0.61591	  &0.65294	  \\
4	&2s  2p$^{2}$	($^{3}$P$^{o}$)	&$^{4}$P	&1.5	&0.64573	   &0.65220	  &0.61705	  &0.65405	  \\
5	&2s  2p$^{2}$	($^{3}$P$^{o}$)	&$^{4}$P	&2.5	&0.64767	   &0.65388	  &0.61860	  &0.65590	  \\
6	&2s  2p$^{2}$	($^{1}$D)	&$^{2}$D	&1.5	 &1.18155	   &1.15686	  &1.20775	  &1.17885  	  \\
7	&2s  2p$^{2}$	($^{1}$D)	&$^{2}$D	&2.5	&1.18156	   &1.15673	  &1.20760	  &1.17890	  \\											
8	&2s  2p$^{2}$	($^{1}$S)	&$^{2}$S	&0.5	&1.48399	   &1.49782	  &1.56308	  &1.53836	  \\											
9	&2s  2p$^{2}$	($^{3}$P$^{o}$)	&$^{2}$P	&0.5	&1.68195	   &1.64466	  &1.73160	  &1.68077	  \\
10	&2s  2p$^{2}$	($^{3}$P$^{o}$)	&$^{2}$P	&1.5	&1.68427	   &1.64688	  &1.73369	  &1.68289	  \\
11	&2p$^{3}$	    ($^{2}$S)	&$^{4}$S$^{o}$	&1.5	&2.11438	   &2.10993	  &2.11408	  &2.11881	  \\
12	&2s$^{2}$ 3d	    ($^{1}$S)	&$^{2}$D	&1.5	&3.96459	   &3.82307	  &3.75789	  &3.83720	  \\
13	&2s$^{2}$ 3d	    ($^{1}$S)	&$^{2}$D	&2.5	&3.96473	   &3.82323	  &3.75803	  &3.83706	  \\
14	&2s 2p 3s	($^{3}$P$^{o}$)	&$^{4}$P$^{o}$	&0.5	&3.98893	   &3.99909	  &3.92915	  &4.00397	  \\
15	&2s 2p 3s	($^{3}$P$^{o}$)	&$^{4}$P$^{o}$	&1.5	&3.99027	   &4.00032	  &3.93038	  &4.00509	  \\
16	&2s 2p 3s	($^{3}$P$^{o}$)	&$^{4}$P$^{o}$	&2.5	&3.99253	   &4.00257	  &3.93261	  &4.00712	  \\
17	&2s 2p 3s	($^{3}$P$^{o}$)	&$^{2}$P$^{o}$	&0.5	&4.14723	   &4.12628	  &4.07957	  &4.14020	  \\
18	&2s 2p 3s	($^{3}$P$^{o}$)	&$^{2}$P$^{o}$	&1.5	&4.14973	   &4.12869	  &4.08194	  &4.14239	  \\
19	&2s 2p 3p	($^{3}$P$^{o}$)	&$^{4}$D	&0.5	&4.24800	   &4.26780	  &4.20150	  &4.27826	  \\											
20	&2s 2p 3p	($^{3}$P$^{o}$)	&$^{4}$D	&1.5	&4.24881	   &4.26851	  &4.20218	  &4.27890	  \\											
21	&2s 2p 3p	($^{3}$P$^{o}$)	&$^{4}$D	&2.5	&4.25016	   &4.26975	  &4.20336	  &4.28001	  \\											
22	&2s 2p 3p	($^{3}$P$^{o}$)	&$^{4}$D	&3.5	&4.25203	   &4.27166	  &4.20526	  &4.28170	  \\											
23	&2s 2p 3p	($^{3}$P$^{o}$)	&$^{4}$S	&1.5	&4.28655	   &4.32376	  &4.25326	  &4.33524	  \\											
24	&2s 2p 3p	($^{3}$P$^{o}$)	&$^{2}$P	&0.5	&4.28710	   &4.25771	  &4.19583	  &4.26653	  \\											
25	&2s 2p 3p	($^{3}$P$^{o}$)	&$^{2}$P	&1.5	&4.28871	   &4.25876	  &4.19691	  &4.26749	  \\											
26	&2s 2p 3p	($^{3}$P$^{o}$)	&$^{4}$P	&0.5	&4.36958	   &4.36359	  &4.29685	  &4.37240	  \\											
27	&2s 2p 3p	($^{3}$P$^{o}$)	&$^{4}$P	&1.5	&4.37041	   &4.36445	  &4.29769	  &4.37323	  \\											
28	&2s 2p 3p	($^{3}$P$^{o}$)	&$^{4}$P	&2.5	&4.37171	   &4.36563	  &4.29882	  &4.37433	  \\											
29	&2s 2p 3p	($^{3}$P$^{o}$)	&$^{2}$D	&1.5	&4.41583	   &4.39838	  &4.35552	  &4.41476	  \\											
30	&2s 2p 3p	($^{3}$P$^{o}$)	&$^{2}$D	&2.5	&4.41823	   &4.40071	  &4.35784	  &4.41681	  \\											
31	&2s 2p 3p	($^{3}$P$^{o}$)	&$^{2}$S	&0.5	&4.49337	   &4.49155	  &4.46991	  &4.52115	  \\											
32	&2s 2p 3d	($^{3}$P$^{o}$)	&$^{4}$F$^{o}$	&1.5	&4.50466	   &4.51231	  &4.45181	  &4.52369	  \\
33	&2s 2p 3d	($^{3}$P$^{o}$)	&$^{4}$F$^{o}$	&2.5	&4.50543	   &4.51302	  &4.45249	  &4.52369	  \\
34	&2s 2p 3d	($^{3}$P$^{o}$)	&$^{4}$F$^{o}$	&3.5	&4.50653	   &4.51405	  &4.45351	  &4.52461	  \\
35	&2s 2p 3d	($^{3}$P$^{o}$)	&$^{4}$F$^{o}$	&4.5	&4.50797	   &4.51545	  &4.45491	  &4.52588	  \\
36	&2s 2p 3d	($^{3}$P$^{o}$)	&$^{4}$D$^{o}$	&0.5	&4.54548	   &4.55421	  &4.48131	  &4.56333	  \\
37	&2s 2p 3d	($^{3}$P$^{o}$)	&$^{4}$D$^{o}$	&1.5	&4.54570	   &4.55447	  &4.48558	  &4.56358	  \\
38	&2s 2p 3d	($^{3}$P$^{o}$)	&$^{4}$D$^{o}$	&2.5	&4.54611	   &4.55490	  &4.48600	  &4.56397	  \\
39	&2s 2p 3d	($^{3}$P$^{o}$)	&$^{4}$D$^{o}$	&3.5	&4.54676	   &4.55549	  &4.48659	  &4.56451	  \\
40	&2s 2p 3d	($^{3}$P$^{o}$)	&$^{2}$D$^{o}$	&2.5	&4.56950	   &4.57059	  &4.51958	  &4.58579	  \\
41	&2s 2p 3d	($^{3}$P$^{o}$)	&$^{2}$D$^{o}$	&1.5	&4.56955	   &4.57009	  &4.51932	  &4.58533  \\											
42	&2s 2p 3d	($^{3}$P$^{o}$)	&$^{4}$P$^{o}$	&2.5	&4.57124	   &4.59366	  &4.52245	  &4.60176	  \\
43	&2s 2p 3d	($^{3}$P$^{o}$)	&$^{4}$P$^{o}$	&1.5	&4.57164	   &4.59469	  &4.52326	  &4.60268	  \\
44	&2s 2p 3d	($^{3}$P$^{o}$)	&$^{4}$P$^{o}$	&0.5	&4.57219	   &4.59536	  &4.52390	  &4.60329	  \\
45	&2s2 4s	    ($^{1}$S)	&$^{2}$S	            &0.5	&4.59636	   &4.42713	  &4.34934	  &4.41291	  \\
46	&2s 2p 3d	($^{3}$P$^{o}$)	&$^{2}$F$^{o}$	&2.5	&4.64907	   &4.65425	  &4.61403	  &	  \\										
47	&2s 2p 3d	($^{3}$P$^{o}$)	&$^{2}$F$^{o}$	&3.5	&4.65133	   &4.65637	  &4.61622	  &	  \\										
48	&2s 2p 3d	($^{3}$P$^{o}$)	&$^{2}$P$^{o}$	&1.5	&4.67580	   &4.68592	  &4.64811	  &	  \\										
49	&2s 2p 3d	($^{3}$P$^{o}$)	&$^{2}$P$^{o}$	&0.5	&4.67709	   &4.68730	  &4.64954	  &4.56333	  \\
50	&2s$^{2}$ 4p	    ($^{1}$S)	&$^{2}$P$^{o}$	&0.5	&4.71223	   &4.55629	  &4.48532	  &	  \\
51	&2s$^{2}$ 4p	    ($^{1}$S)	&$^{2}$P$^{o}$	&1.5	&4.71258	   &4.55668	  &4.48168	  &4.56407	  \\
52	&2s 2p 3s	($^{1}$P$^{o}$)	&$^{2}$P$^{o}$	&0.5	&4.78493	   &4.72673	  &4.75301	  &	  \\										
53	&2s 2p 3s	($^{1}$P$^{o}$)	&$^{2}$P$^{o}$	&1.5	&4.78501	   &4.72682	  &4.75313	  &	  \\										
54	&2s$^{2}$ 4d	    ($^{1}$S)	&$^{2}$D	&1.5	&4.81167	   &4.65265	  &4.57274	  &	  \\
55	&2s$^{2}$ 4d	    ($^{1}$S)	&$^{2}$D	&2.5	&4.81171	   &4.65269	  &4.57279	  &	  \\
56	&2s$^{2}$ 4f	    ($^{1}$S)	&$^{2}$F$^{o}$	&2.5	&4.82628	   &4.67651	  &4.59047	  &	  \\
57	&2s$^{2}$ 4f	    ($^{1}$S)	&$^{2}$F$^{o}$	&3.5	&4.82631	   &4.67661	  &4.59051	  &	  \\
58	&2s 2p 3p	($^{1}$P$^{o}$)	&$^{2}$D	&1.5	&5.04651	   &4.98760	  &5.03332	  &  \\										
59	&2s 2p 3p	($^{1}$P$^{o}$)	&$^{2}$D	&2.5	&5.04675	   &4.98787	  &5.03353	  &  \\										
60	&2s 2p 3p	($^{1}$P$^{o}$)	&$^{2}$P	&0.5	&5.07598	   &5.01007	  &5.03586	  &  \\										
61	&2s 2p 3p	($^{1}$P$^{o}$)	&$^{2}$P	&1.5	&5.07660	   &5.01065	  &5.03645	  &  \\										
62	&2s 2p 3p	($^{1}$P$^{o}$)	&$^{2}$S	&0.5	&5.14024	   &5.05265	  &5.11693	  &	  \\										
63	&2s 2p 3d	($^{1}$P$^{o}$)	&$^{2}$D$^{o}$	&1.5	&5.28047	   &5.24725	  &5.30203	  &  \\										
64	&2s 2p 3d	($^{1}$P$^{o}$)	&$^{2}$D$^{o}$	&2.5	&5.28067	   &5.24756	  &5.30232	  &  \\										
65	&2s 2p 3d	($^{1}$P$^{o}$)	&$^{2}$F$^{o}$	&3.5	&5.29185	   &5.20148	  &5.27744	  &  \\										
66	&2s 2p 3d	($^{1}$P$^{o}$)	&$^{2}$F$^{o}$	&2.5	&5.29194	   &5.20143	  &5.27749	  &  \\										
67	&2s 2p 3d	($^{1}$P$^{o}$)	&$^{2}$P$^{o}$	&0.5	&5.35379	   &5.30103	  &5.35907	  &  \\										
68	&2s 2p 3d	($^{1}$P$^{o}$)	&$^{2}$P$^{o}$	&1.5	&5.35391	   &5.30123	  &5.35929	  & \\										

    \hline
  \end{tabular}
 }
\end{table}

\begin{onecolumn}
\begin{table}
  \centering
  \caption{Comparison of our values of log$(g_{i}f_{ij})$ and transition probability $A_{ji} ($s$^{ - 1} )$ with the values of \cite{Galavis98}, \cite{Flower75} and \cite{Tayal06} for 5 intercombination transitions. Observed wavelengths are from \cite{Bromander69}.}\label{Tab2}
  \resizebox{14cm}{!} {
  \begin{tabular}{cclccccrrrrr}
    \hline
Index	&Transition	 &wav (\AA )	&Conf${_i}$		         &J${_i}$	   &Conf${_f}$		    &J${_f}$ &log(gf)	 &A(s$^{-1}$)	    &A(Galavis)	 &A(Flower)	 &A(Tayal)\\
\hline
1	&1 $-$ 4	&1397.2	    &2s$^{2}$ 2p	$^{2}$P$^{o}$  	&0.5	&2s  2p$^{2}$	$^{4}$P	&1.5	&-7.316	&40.69	    &48.36	    &58.2	    &36.6   \\
2	&1 $-$ 3	&1399.77	&2s$^{2}$ 2p	$^{2}$P$^{o}$	&0.5	&2s  2p$^{2}$	$^{4}$P	&0.5	&-5.98	&1757.58	 &1724	    &2080	    &1280   \\
3	&2 $-$ 5	&1401.16	&2s$^{2}$ 2p	$^{2}$P$^{o}$	&1.5	&2s  2p$^{2}$	$^{4}$P	&2.5	&-5.688	&1146.37	 &1331	    &1470	    &1025   \\
4	&2 $-$ 4	&1404.81	&2s$^{2}$ 2p	$^{2}$P$^{o}$	&1.5	&2s  2p$^{2}$	$^{4}$P	&1.5	&-6.347	&374.79	    &314.7	    &441	    &251.3  \\
5	&2 $-$ 3	&1407.39	&2s$^{2}$ 2p	$^{2}$P$^{o}$	&1.5	&2s  2p$^{2}$	$^{4}$P	&0.5	&-5.973	&1766.95	 &1762	    &2130	    &1264   \\
    \hline
  \end{tabular}
 }
\end{table}
\end{onecolumn}

\begin{twocolumn}
\begin{figure}
\label{figure1}
\begin{center}
\includegraphics*[width=10cm]{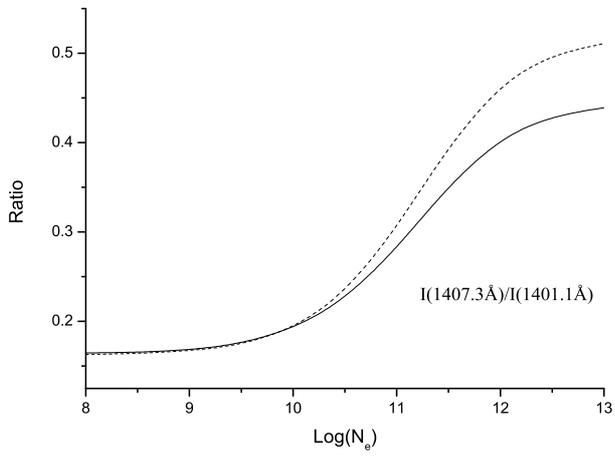}
\end{center}
\caption{Our calculations: line ratio I(1407.3 \AA )/I(1401.1 \AA ) plotted as a function of logarithmic electron density (N$_e$ in cm$^{-3}$) and at a constant electron temperature of T$_e$=10$^5$ K. The solid curve is obtained by using the oscillator strengths from \cite{Galavis98}, and the dashed curve is obtained by using our HFR calculations for the 5 intercombination transitions.}
\end{figure}
\end{twocolumn}
\end{document}